# When Does Your Brain Know You? Segment Length and Its Impact on EEG-based Biometric Authentication Accuracy.

Nibras Abo Alzahab [*1], Lorenzo Scalisae[1]., Marco Baldi[3]

[1] Marche Polytechnic University (UNIVPM), N.Abo_Alzahab@pm.univpm.it.

[2] Marche Polytechnic University (UNIVPM), L.Scalise@univpm.it.

[3] Marche Polytechnic University (UNIVPM), M.Baldi@univpm.it.

* corresponding authors.

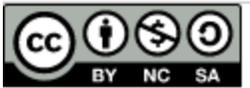

## Abstract:

In the quest for optimal EEG-based biometric authentication, this study investigates the pivotal balance for accurate identification without sacrificing performance or adding unnecessary computational complexity. Through a methodical exploration of segment durations, and employing a variety of sophisticated machine learning models, the research seeks to pinpoint a threshold where EEG data provides maximum informational yield for authentication purposes. The findings are set to advance the field of non-invasive biometric technologies, proposing a practical approach to secure and user-friendly identity verification systems while also raising considerations for the real-world application of EEG-based biometric authentication beyond controlled environments.

## 1. Introduction:

Neuroscience, the study of the nervous system, has paved the way for groundbreaking advancements in understanding the brain's intricate processes and functions (Abbasi & Rizzo, 2021). Among these advancements, Brain-Computer Interfaces (BCIs) stand out as a revolutionary technology that bridges the gap between the human brain and external devices, enabling direct communication without physical movement (Abo Alzahab et al., 2019; Nicolas-Alonso & Gomez-Gil, 2012; Wolpaw et al., 2000a, 2000b). Specifically, in the realm of security and authentication, BCIs have introduced a novel biometric modality that leverages the unique neural patterns of individuals. Electroencephalography (EEG), a method for recording electrical activity of the brain, has emerged as a prominent technique in this context (Abo Alzahab et al., 2022). By analyzing EEG signals, which are distinct and difficult to replicate, it is possible to create highly secure and personalized biometric authentication systems. This approach not only enhances security but also offers a seamless, user-friendly method of verifying identity, showcasing the potential of integrating neuroscience with cybersecurity to redefine traditional security measures (Bak & Jeong, 2023).

In the ever-evolving landscape of biomedical engineering, the integration of Brain-Computer Interfaces (BCIs) with biometrics represents a pioneering frontier that heralds a new era of secure and personalized technology (Paranjape et al., n.d.). BCIs, at their core, facilitate a direct communication pathway between the brain and external devices, leveraging neural signals to control or interact with computers and other machinery (Abo Alzahab et al., 2022; Das et al., 2016). This remarkable capability has profound implications for medical signal processing, particularly when applied to the realm of biometrics, where the unique patterns of brain activity offer an unparalleled level of security and individual identification (Wang et al., 2019). The novelty of this research lies in its exploration of Electroencephalography (EEG) biometrics, a domain where the electrical activity of the brain is mined for data that can uniquely identify individuals across various cognitive states and conditions.

The crux of this research lies in delving deeper into the utilization of EEG signals for authentication purposes, focusing particularly on identifying a temporal threshold that encapsulates the minimum amount of information required for effective authentication. This inquiry starts from the assumption that shorter EEG segments may contain an insufficient amount of information, rendering them ineffective for reliable identity verification. Conversely, excessively long EEG segments can lead to increased computational complexity and latency, as well as reduced user convenience, thus straining the practical application of this technology in real-world scenarios. The novelty of this research stems from its attempt to find an optimal balance between these two extremes, thereby pioneering a more efficient, accurate, and user-friendly EEG-based biometric authentication system. By exploring this temporal threshold, this study aims to push the boundaries of current knowledge and open new avenues for non-invasive biometric security technologies. The importance of this research cannot be overstated, as it holds the potential to improve the way we approach security and identity verification, offering a solution that is both highly secure against fraudulent attempts and adaptable to the dynamic needs of users in various contexts. Through this work, we envision a future where the unique electrical patterns of our brain become the key to unlocking a new dimension of personal security and technological interaction.

## 2. Literature Review:

The exploration of EEG biometrics has diversified, focusing on segment duration, methodological innovations, and applications in user recognition and security. This section reviews pivotal studies that contribute to understanding the optimal EEG segment duration and methodological advancements in EEG-based biometric identification.

Authors in (Das et al., 2015) investigated the efficacy of varied intervals post-stimulus in EEG signal analysis for biometric recognition. They have tested all the possible time intervals between {0; 100; 200; 300; 400; 500; 600; 700} ms. They have found that the usage of full-length interval gives the best performance. They have not considered any interval greater than 0.700 seconds.

The paper by (Bak & Jeong, 2023) presents an EEG-MI methodology utilizing optimized feature extraction methods and classifiers to enhance user identification accuracy. It explores the efficacy of four features (CSP, ERD/S, AR, FFT) with SVM and GNB classifiers using 4-seconds EEG segment, achieving high accuracies (up to 98.97% with SVM) and validating results with the half-total error rate (HTER) to ensure reliability despite data size imbalances. The study significantly contributes to information security in biometric identification, demonstrating the potential of EEG-MI signals for accurate user recognition.

The research conducted by (Carrión-Ojeda et al., 2019) has employed Discrete Wavelet Transform and a greedy strategy for hyperparameter selection, with findings pointing to 2 seconds for achieving 90% accuracy. This study underpins the efficiency of short-duration EEG recordings in biometric identification, aligning closely with our investigation into temporal thresholds.

While authors in (Ozdenizci et al., 2019) leverage half-second EEG epochs, utilizing a convolutional neural network with an adversarial component to improve cross-session identification accuracy. This study's focus on short epochs (0.5 seconds) contributes to the target of optimizing EEG segment length for biometric purposes, emphasizing the potential of adversarial learning in enhancing system robustness.

The work (Oike-Akino et al., 2016) showcases the use of ERP epochs and dimensionality reduction techniques (PCA/PLS) alongside machine learning classifiers. With a segment length of 0.8 seconds for identification, this research highlights the effectiveness of combining dimensionality reduction with machine learning to achieve high identification accuracy, further supporting the relevance of short-duration EEG segments in biometric applications.

On the other hand, some publications do not specify EEG segment durations; they contribute significantly to the field through enhanced data security, signal processing advancements, and user identification techniques. Particularly, the integration of EEG biometrics with cryptographic techniques and the signal acquisition and classification underscore the evolving landscape of EEG biometric security and its application in emerging technologies like 5G and IoT (Beyrouthy et al., 2024; Gui et al., 2014, 2015).

In conclusion, these studies, summarized in Table 1, underscore the critical role of EEG segment duration in optimizing biometric accuracy and efficiency. The methodological innovations presented further enrich the field, providing a solid foundation upon which our research builds to explore the temporal threshold for EEG-based biometrics.

**Table 1 – Overview of related work**

| Reference | Methodology | EEG segment length in seconds | Contributions to EEG Biometrics |
|---|---|---|---|
| (Bak & Jeong, 2023) | Motor Imagery Dataset, feature extraction, and classification | 4 seconds | Achieving high performance using MI-based EEG signals. |
| (Carrión-Ojeda et al., 2019) | Discrete Wavelet Transform and greedy strategy for hyperparameter selection | 2 seconds | Demonstrated the efficiency of short-duration EEG recordings for biometric identification |
| (Ozdenizci et al., 2019) | Convolutional neural network with adversarial component | 0.5 seconds | Improved cross-session identification accuracy using adversarial learning |
| (Damaševičius et al., 2018) | Integration of EEG biometrics with cryptographic techniques | 4 seconds | Enhanced data security in EEG-based biometric authentication |
| (oike-Akino et al., 2016) | PCA/PLS and machine learning classifiers | 0.8 seconds | Highlighted the effectiveness of dimensionality reduction techniques |
| (Das et al., 2015) | Analysis of EEG signals for biometric recognition | 0.7 seconds | Identified the most informative EEG sub-band and effective time interval |
| (Gui et al., 2015) | Euclidean Distance and Dynamic Time Warping | 1.1 seconds | Enhanced user identification through specific stimuli and channels |
| (Gui et al., 2014) | Signal acquisition, noise reduction, feature extraction, and classification | 1.1 seconds | Enhanced user identification and authentication accuracy through processing techniques |

## 3. Material and Methods:

In this section, we describe the methods used to determine the temporal threshold we are looking for. First, we discuss the setup of our experiments. Then, we provide an overview of the machine learning algorithm used and the training process.

### 3.1 Experimental Setup

Our experimental setup includes a detailed description of the datasets used in our research. We provide a general overview of each dataset and explain the scenarios we followed for segmenting EEG signals based on temporal intervals. Finally, we list all the features extracted that demonstrate the variations in identity shown by EEG signals.

#### 3.1.1 Dataset Description

STEW Dataset: Selected for its inclusion of both mental workload and resting state scenarios, the STEW dataset allows for the examination of EEG biometric performance under varying cognitive states. This variability is crucial for assessing the robustness of EEG biometrics across different user states, mirroring real-world applications where users might interact with biometric systems under various cognitive loads.

The Simultaneous Task EEG Workload Data Set (STEW) constitutes a pivotal resource for the examination of brain dynamics across mental workload and resting states, documented through electroencephalography (EEG). This dataset, meticulously compiled from 48 individuals, bifurcates into two distinct segments: one capturing the neural quietude of resting states and the other detailing the cerebral exertion inherent to mental workload scenarios, facilitated by engagement with the SIMKAP multitasking test.

Encompassing EEG recordings spanning 2.5 minutes across 14 strategically chosen channels (AF3, F7, F3, FC5, T7, P7, O1, O2, P8, T8, FC6, F4, F8, and AF4), this dataset provides a large amount of information concerning brain activity, sampled at a frequency of 128 Hz. Preliminary data processing was conducted using a first-order bandpass Butterworth filter, with a frequency range of 3 to 40 Hz, to refine the signals by isolating the spectrum of brain wave frequencies pertinent to cognitive load analysis. The STEW dataset emerges as an invaluable asset for advancing research within cognitive neuroscience and psychology, offering nuanced insights into the neural activity correlated with the of cognitive load and tranquility.

EEG Alpha Wave Dataset: Chosen for its focus on alpha wave patterns, which are prominent during relaxed, calm states. This dataset is instrumental in understanding how well EEG biometric systems can identify individuals based on brainwave patterns associated with relaxation, a common state for many users.

The EEG Alpha Waves Dataset is a specialized repository of EEG signals that encapsulate the neural state of repose, gathered from 20 participants. This dataset provides 2 minutes of EEG recordings, obtained from a comprehensive array of 16 channels (FP1, FP2, FC5, FC6, FZ, T7, CZ, T8, P7, P3, PZ, P4, P8, O1, Oz, and O2), and showcases a higher sampling rate of 512 Hz. To ensure the fidelity and analytical relevance of the data, a preprocessing step was applied, employing a first-order bandpass Butterworth filter within the 3 - 40 Hz frequency range. This dataset, therefore, stands as a critical resource for the exploration of resting state neural dynamics, underpinning studies aimed at deciphering the complexity of brain function in tranquility.

### 3.1.2 Segmentation

The selection of 19 different segment lengths, ranging from 0.1 to 10 seconds, is for identifying the optimal temporal threshold for EEG biometric systems, as illustrated in Fig. 1. These varying lengths allow for a comprehensive analysis of how the duration of EEG recording impacts the accuracy and reliability of biometric identification. Shorter segments may offer convenience and speed, while longer segments could improve accuracy, but at the cost of increased processing time and potentially reduced user comfort. This range enables a detailed exploration of the trade-off between speed and accuracy in EEG biometric systems.

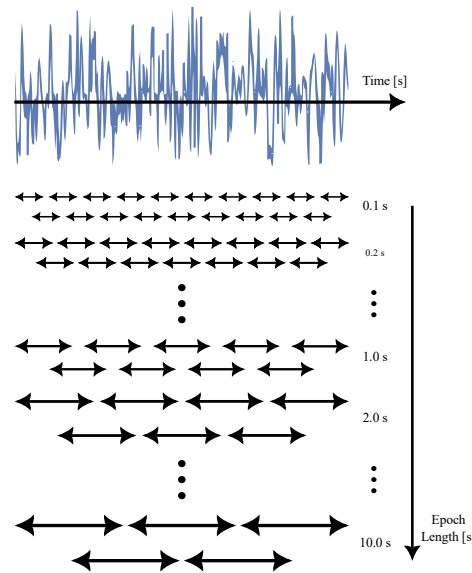

Figure 1 – EEG-Segmentation Scenarios.

### 3.1.3 Feature Extraction

Based on empirical analysis, the study rigorously examined 19 features for the purpose of feature selection. These features are systematically organized into five distinct categories, as delineated below:

1. Statistical Features:
   a. Mean Amplitude
   b. Standard Deviation of Amplitude
   c. Variance of Amplitude
   d. Range of Amplitude (Peak-to-Peak)
   e. Skewness of Amplitude
   f. Kurtosis of Amplitude
2. Power Spectral Density (PSD) Features:

a. Power in the Delta Band
   b. Power in the Theta Band
   c. Power in the Alpha Band
   d. Power in the Beta Band
   e. Power in the Gamma Band
3. Entropy Features:
   a. Permutation Entropy
   b. Entropy Derived from Singular Value Decomposition (SVD)
   c. Approximate Entropy
   d. Sample Entropy
4. Fractional Dimension (FD) Features:
   a. Petrosian Fractal Dimension
   b. Katz Fractal Dimension
   c. Higuchi Fractal Dimension
5. Detrended Fluctuation Analysis (DFA)

For an in-depth description of these features and their implications for EEG biometric analysis, refer to the comprehensive descriptions provided in the literature (Abo Alzahab, 2021).

### 3.2 Classification Techniques

This section gives a general overview of the classification algorithms used in the experimental work. Those are Multilayer perceptron (MLP), K-Nearest Neighbors, and eXtreme gradient Boosting (XGBoost)

#### 3.2.1 Multilayer Perceptron (MLP)

The MLP architecture was tailored for each dataset through a trial and error approach to optimise performance. The architecture for the STEW dataset consists of 5 dense layers with a specific arrangement of neurons to handle the complexity of mental workload and resting state data. In contrast, the architecture for the EEG Alpha dataset is somewhat simpler, reflecting the focused nature of this dataset on alpha wave patterns. The selection of activation functions (ReLU and Softmax), the optimizer (Adam), batch size, and training duration (1000 epochs) were all chosen to balance the trade-off between training efficiency and model performance. The details of the MLP are presented in Table 1. The general diagram of the evaluation of the authentication performance is presented in Table 2.

**Table 2 – Multilayer Perceptron layers**

|  | STEW Dataset | | EEG Alpha Wave Dataset | |
| --- | --- | --- | --- | --- |
| **Layers** | Neurons | Activation Function | Neurons | Activation Function |
| **Dense 1** | 200 | ReLU | 200 | ReLU |
| **Dense 2** | 13150 | ReLU | 120 | ReLU |
| **Dense 3** | 100 | ReLU | 70 | ReLU |
| **Dense 4** | 75 | ReLU | 19 | Softmax |
| **Dense 5** | 48 | Softmax | | |

### 3.2.2 K-Nearest Neighbors (KNN)

KNN is a straightforward classifier that relies on majority voting from the nearest data points. The experimentation with different numbers of neighbors aims to fine-tune the model for optimal simplicity and effectiveness.

### 3.2.3 eXtreme Gradient Boosting (XGBoost)

This advanced ensemble method enhances performance by combining simpler models. It is recognized for its high efficiency and low computational demands, making it a valuable tool for handling complex EEG data.

### 3.3 Training

The training process included testing each algorithm with various feature sets and segment durations, as well as under different experimental conditions. To mitigate the effects of classifier randomness and ensure the robustness of the findings, the classification was repeated three times, with performance metrics averaged and standard deviation recorded.

## 4. Results

This study adopted 19 segmentation scenarios to ascertain a temporal threshold indicative of the highest information content for EEG signals. The search for this temporal threshold is critical, as it promises to enhance the efficiency of biometric identification processes significantly.

Fig 2 presents the normalised accuracies for the three different machine learning models: Multilayer Perceptron (MLP), K-Nearest Neighbors (KNN), and eXtreme Gradient Boosting (XGB). Across the duration spectrum, each model exhibits a distinct behaviour pattern in terms of accuracy. Notably, for durations longer than 2 seconds, the accuracy rates reach a plateau, suggesting a convergence towards maximum information extraction from the EEG signals.

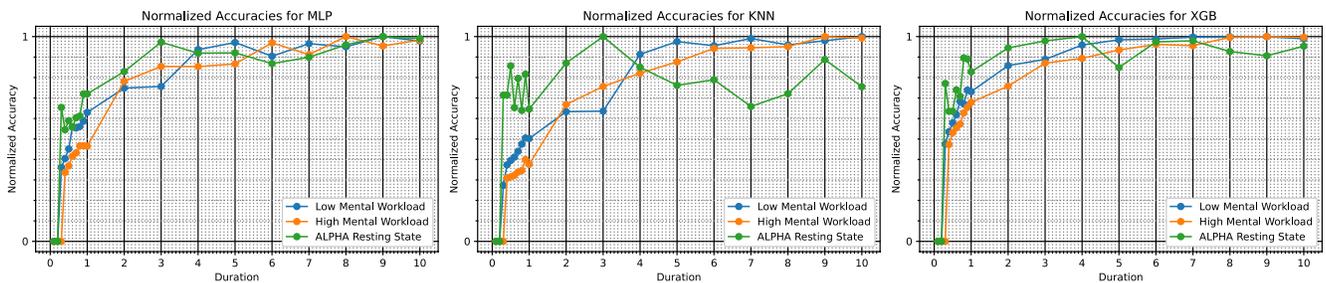

Figure 2 – Accuracy of EEG-based biometric authentication using MLP, KNN, and XGB algorithms, across various segment durations.

Considering the unique features and foundational assumptions of each model is essential, as these factors significantly impact their performance. For instance

- MLP: Demonstrates a sharp increase in normalized accuracy within the initial second, followed by a plateau that indicates a steady state of accuracy.
- KNN: Shows a more variable pattern, with notable fluctuations in accuracy, yet still approaching a plateau post the 2-second mark.
- XGB: Exhibits a consistent ascent to a plateau of accuracy, maintaining a steady state beyond the elbow point identified at 2 seconds.

Fig 3 reports the derivative of normalized accuracies for the MLP, KNN, and XGB models. These derivatives serve as a lens into the rate of change in accuracy, offering insights into the dynamics of information gain over time.

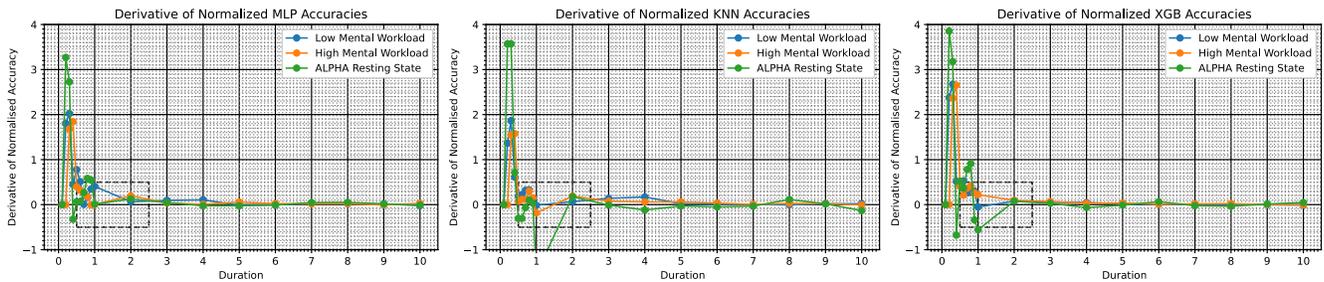

Figure 3 – Elbow Diagram Highlighting the Derivatives of Normalized Accuracy Over Segment Duration for MLP, KNN, and XGB Classifiers. The dashed box pinpointing the critical elbow threshold where the accuracy improvements plateau, suggesting an optimal EEG segment duration for authentication purposes.

It is crucial to account for the distinct characteristics and underlying principles of each model, given their substantial influence on performance outcomes.

- MLP: The early derivative peaks suggest a rapid initial gain in information, which stabilizes quickly as evidenced by the derivative approaching zero, correlating with the accuracy plateau observed previously.
- KNN: Exhibits a more turbulent derivative profile, with multiple peaks indicating variable information gain rates. However, the general trend moves towards stabilization, especially beyond the 2-second threshold.
- XGB: The change in accuracy remains comparatively consistent, with a single peak followed by a gradual decline towards zero, reinforcing the notion of an optimal information gain threshold.

Identifying the elbow value at precisely 2 seconds has been a pivotal aspect of this analysis. Employing techniques such as the "knee point detection method" referenced in prior studies (Satopää et al., 2011), this temporal marker identifies the point at which additional duration contributes minimally to the enhancement of identification accuracy. This finding is particularly significant for EEG biometrics, as it suggests a reduction in data processing time without compromising the accuracy of biometric identification.

When compared with existing literature, the temporal threshold of 2 seconds reinforces findings from similar studies (Carrión-Ojeda et al., 2019; Gómez-Tapia et al., 2022), which have identified thresholds within the same range as seen in Fig 4. Additionally, a correlation factor equal to 0.97 and 0.94 with P-Value equals to 6.48e-5 and 4.96e-5 was found between our results and (Carrión-Ojeda et al., 2019) and (Gómez-Tapia et al., 2022), respectively. This study's confirmation of the 2-second elbow value contributes to the consensus on this topic, while also enhancing the methodological robustness through the application of multiple advanced machine learning models.

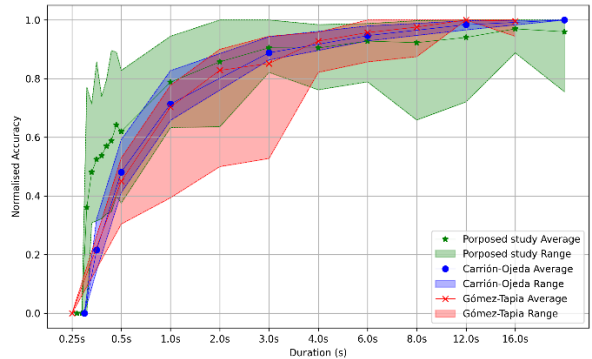

Figure 4 – Comparison of the proposed study with literature.

The investigations detailed herein have culminated in the affirmation of a 2-second temporal threshold for EEG-based biometrics. This threshold delineates a balance between the duration of EEG signal segments and the fidelity of biometric identification, providing a benchmark for future applications in the domain.

Setting a time limit has many benefits. First, it makes identifying people using biometrics quicker and more accurate by needing less data. Also, it can help make the EEG data acquisition paradigms more consistent, making it easier to towards standardized EEG-based biometrics.

## 5. Discussion

Through this work, EEG signals were used to determine the optimal temporal threshold for EEG signals. This was achieved by using different EEG data sets and different machine learning algorithms. The EEG signals were segmented into 19 scenarios withing the range [0.1-10] seconds. This study's establishment of a 2-second temporal threshold for EEG signal segmentation has significant ramifications for the field of EEG biometrics. Not only does this finding streamline the signal processing stage, but it also aligns closely with the temporal resolutions reported in contemporary literature (Carrión-Ojeda et al., 2019). The coherence of these findings suggests a strong reliability in the 2-second threshold as a staple for EEG-based identification systems.

The differential performance of the MLP, KNN, and XGB models provides insightful revelations into the suitability of various machine learning algorithms for EEG data analysis. While MLP and XGB models show a relatively stable accuracy after the 2-second mark, the KNN model's fluctuating accuracy underscores its sensitivity to the choice of k and the distance metric used. This sensitivity could be leveraged in systems where dynamic responsiveness to signal variations is paramount.

The delineation of a 2-second EEG signal segmentation threshold holds promise for substantially enhancing the efficiency of biometric identification systems. This optimized duration could lead to quicker authentication processes, reduced computational costs, and potentially, the inclusion of EEG biometrics in a broader array of applications, including secure access control and user authentication in computing systems.

While the findings are robust and go in line with existing literature, the study is not without limitations. The primary limitation is the reliance on specific machine learning models, which may exhibit varied performance across different datasets or EEG acquisition protocols. Furthermore, the study's scope was confined to a controlled environment, which might not encapsulate the complexity of real-world settings.